%% file: revised_r2.tex
\documentclass[journal]{IEEEtran}
\IEEEoverridecommandlockouts

\usepackage{amsthm,amssymb,graphicx,multirow,amsmath,color,amsfonts}
\usepackage[update,prepend]{epstopdf}
\usepackage[noadjust]{cite}
\usepackage[latin1]{inputenc}
\usepackage{tikz}
\usetikzlibrary{arrows,calc}		
\usepackage{bbm} 
\usepackage{pdfpages}
\usepackage{tabulary}
\usepackage{multirow}
\usepackage{comment}
\usepackage{subcaption} 
\usepackage{caption}  
\usepackage[ruled,vlined,linesnumbered]{algorithm2e}
\pdfoutput=1
\usepackage{svg}
\usepackage{enumitem,kantlipsum}
\usepackage{graphicx} 
\usepackage{balance}
\def\chb#1{{\color{blue}#1}}

\renewcommand{\chb}[1]{\textcolor{black}{#1}}

\include{notation}

\allowdisplaybreaks 

\begin{document}
\graphicspath{{./Figures/}}
\title{
LEO-based Positioning: Foundations, Signal Design, and Receiver Enhancements for 6G NTN}
\author{
Harish K. Dureppagari*, Chiranjib Saha*, Harikumar Krishnamurthy, Xiao Feng Wang, Alberto Rico-Alvari\~{n}o, R. Michael Buehrer, Harpreet S. Dhillon
\thanks{H. K. Dureppagari, R. M. Buehrer, and H. S. Dhillon are with Wireless@VT, Department of ECE, Virginia Tech, Blacksburg, VA 24061, USA. Email: \{harishkumard, rbuehrer, hdhillon\}@vt.edu. C. Saha, H. Krishnamurthy, X. F. Wang, and A. Rico Alvari\~{n}o are with the Qualcomm Standards and Industry Organization, Qualcomm Technologies Inc., San Diego, CA 92121, USA. Email: \{csaha, harkris, wangxiao, albertor\}@qti.qualcomm.com. The support of the US NSF (Grants CNS-1923807 and CNS-2107276) is gratefully acknowledged.

* These authors contributed equally. 
}
\vspace{-6mm}
}

\maketitle

\begin{abstract}
The integration of non-terrestrial networks (NTN) into 5G new radio (NR) has opened up the possibility of developing a new positioning infrastructure using  NR signals from Low-Earth Orbit (LEO) satellites. \chb{Compared to existing Global Navigation Satellite Systems (GNSS),} LEO-based cellular positioning offers several advantages, such as a superior link budget, higher operating bandwidth, and large forthcoming constellations. Due to these factors, LEO-based positioning, navigation, and timing (PNT) is a potential enhancement for NTN in 6G cellular networks. However, extending the existing terrestrial cellular positioning methods to LEO-based NTN positioning requires key fundamental enhancements. These include creating broad positioning beams orthogonal to conventional communication beams, time-domain processing at the user equipment (UE) to resolve large delay and Doppler uncertainties, and efficiently accommodating positioning reference signals (PRS) from multiple satellites within the communication resource grid. In this paper, we present the first set of design insights by incorporating these enhancements and thoroughly evaluating LEO-based positioning, considering the constraints and capabilities of the NR-NTN physical layer. To evaluate the performance of LEO-based NTN positioning, we develop a comprehensive NR-compliant simulation framework, including LEO orbit simulation, transmission (Tx) and receiver (Rx) architectures, and a positioning engine incorporating the necessary enhancements. Our findings suggest that LEO-based NTN positioning could serve as a complementary infrastructure to GNSS and, with appropriate enhancements, may also offer a viable alternative.



\end{abstract}

\begin{IEEEkeywords}
NTN positioning, LEO-based positioning, LEO constellation, coarse positioning, precise positioning.
\end{IEEEkeywords}
\vspace{-1mm}
\section{Introduction}\label{sec:intro}
Positioning has been an essential component in cellular services, initially driven by regulatory requirements for determining the location of emergency calls. The need for precise positioning has grown significantly with the proliferation of navigation-dependent applications and advanced use cases such as smart factories, autonomous driving, and augmented and virtual reality (AR/VR), especially as we move towards 6G networks. In the 5G era, cellular positioning has evolved significantly with advancements, including assisted GNSS, downlink (DL) and uplink (UL) positioning methods, multi-technology integration (such as Bluetooth and WLAN), and sidelink positioning. Starting from Release 17, NTN was introduced in 5G NR, supporting cellular services through satellites in low-earth orbits (LEO), medium-earth orbits (MEO), and geostationary-earth orbits (GEO)~\cite{dureppagari_ntn_10355106,3gpp::38300}. This paves the way for emerging satellite constellations (e.g., Starlink and OneWeb) to provide enhanced mobile broadband (EMBB) and Internet-of-things (IoT) services globally. 

Although GNSS will remain a cornerstone for PNT, it is crucial to understand if LEO constellations can offer a promising alternative for PNT using 6G cellular infrastructure, which is the main theme of this paper. \chb{LEO satellites, orbiting at around 600 km, offer superior link budgets, enhancing signal strength compared to GNSS, which orbit at 20,200 km. Improved signal strength, coupled with larger bandwidth, improves multipath resolution and reduces time-to-first-fix (TTFF) compared to GNSS}. Additionally, LEO-based systems can eliminate the need for GNSS radios (as of Release 19, NR-NTN connectivity requires UE to resolve its location using GNSS), reducing UE power consumption and complexity. However, compared to GNSS, repurposing communication-focused LEO satellites for positioning poses several challenges: narrow and frequency-separated beams are less suitable for positioning compared to the broader beams in GNSS, limited phase coherence hinders coherent combining of signals across time, and high variability in delay and Doppler shifts adds additional complexity in detecting and tracking PRS signals.

\begin{figure*}
    \centering
    \includegraphics[width=0.90\linewidth]{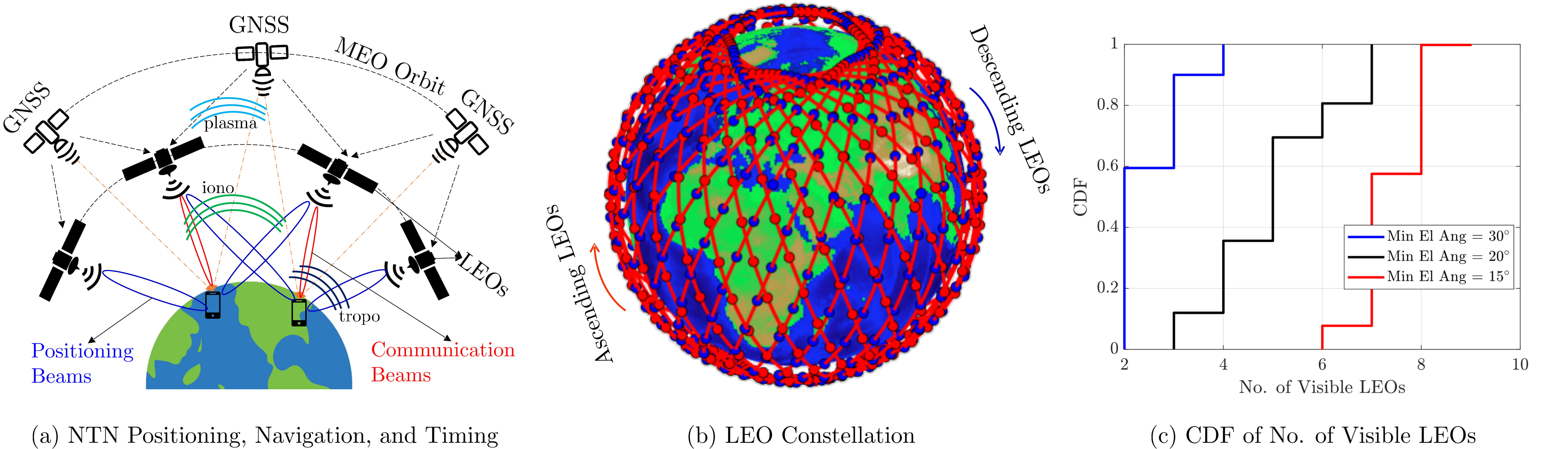}
    \captionsetup{font=small}
    \caption{(a) NTN positioning, navigation, and timing. (b) LEO constellation. (c) CDF of the number of visible LEOs considering UE drop near the equator (worst case scenario).}
    \label{fig:ntn_pnt}
    \vspace{-16pt} 
\end{figure*}

{\em Prior Art.}
The problem of understanding the localization capability of satellite-based communication services is not new. For instance, in~\cite{levanon_705883}, the authors proposed a positioning method for the Globalstar system based on time and frequency differences of arrival (TDOA/FDOA) using one or a few LEO satellites. However, the problem setup fundamentally changes when we try to design a similar positioning system integrated into NR-NTN, which is the focus of this article. As of Release 19, unlike terrestrial networks (TNs), positioning using the NTN infrastructure is not supported~\cite{3gpp::23273}. Although single-satellite-based coarse positioning enables UE location verification within 10 km~\cite{dureppagari_ntn_10355106}, multi-satellite-based precise positioning remains out of scope for NTN in Release 19.

To put the contribution of this paper into perspective, we point out the difference between this work and the volume of recent works on LEO-based positioning, where the DL signal from LEO is used as a {\em signal of opportunity}~\cite{zak_10605899,9840374}. These approaches require a complex receiver setup to be able to track the phase of the received signals coherently for a very long time (coherent processing time of 200 ms assumed in~\cite{zak_10605899}), assuming no knowledge of the transmitted signals, hence requiring high receiver sensitivity, and assume LEO satellites to be phase coherent during measurement duration. Additionally, recent research has explored interference analysis for PRS transmission in 5G NTN in the context of TN/NTN integration~\cite{10140024,10759698}, reusing the TN PRS framework. \chb{While some efforts are ongoing to support standalone LEO-PNT frameworks (e.g., Xona) and GNSS augmentation using LEO satellites (e.g., Centispace)~\cite{_elikbilek_2024}, our work focuses on leveraging communication-centric LEO constellations for positioning rather than developing a dedicated LEO-PNT system.} In this work, we propose design enhancements for NR-NTN to achieve positioning performance comparable to GNSS with small TTIF at NR UEs (e.g., smartphones) with limited hardware availability. 
Building upon our previous work~\cite{dureppagari_ntn_10355106}, where we examined NTN-positioning study cases for 6G using Cram\'er-Rao lower bound (CRLB)-based analysis, this paper provides the first set of design guidelines for a LEO-based positioning system utilizing NR-NTN cellular services.


{\em Contributions.} To assess the feasibility of LEO-based NTN positioning as a complementary infrastructure to GNSS, we evaluate the positioning performance considering communication-focused LEO constellations and NR-NTN physical layer aspects. We begin by analyzing the feasibility of such constellations for positioning by examining the number of visible LEOs at various elevation angles. Next, we provide signal design guidelines to assist UEs in acquiring PRS and estimating delay and Doppler from multiple satellites within a measurement window. We then provide the outline of a {\em simulation framework} to evaluate the performance of LEO-based NTN positioning. To this end, we introduce the methods for signal generation, focusing on accurately modeling fundamental aspects of LEO-based systems, including time-varying delay and Doppler shifts. Following this, we outline the receiver architecture, including algorithms for acquiring the PRS, estimating the initial delay (coarse TDOA) and Doppler, and computing the coarse UE location. To further enhance accuracy, we explore combining delay measurements over time and employing multi-symbol PRS transmissions, demonstrating how this can yield positioning performance comparable to GNSS. We conclude our discussion by suggesting that NTNs hold significant potential for positioning, and further performance improvements can be achieved through enhanced PRS signal transmission, reception, and large constellations.

\begin{figure*}
    \centering
    \includegraphics[width= 0.74\linewidth]{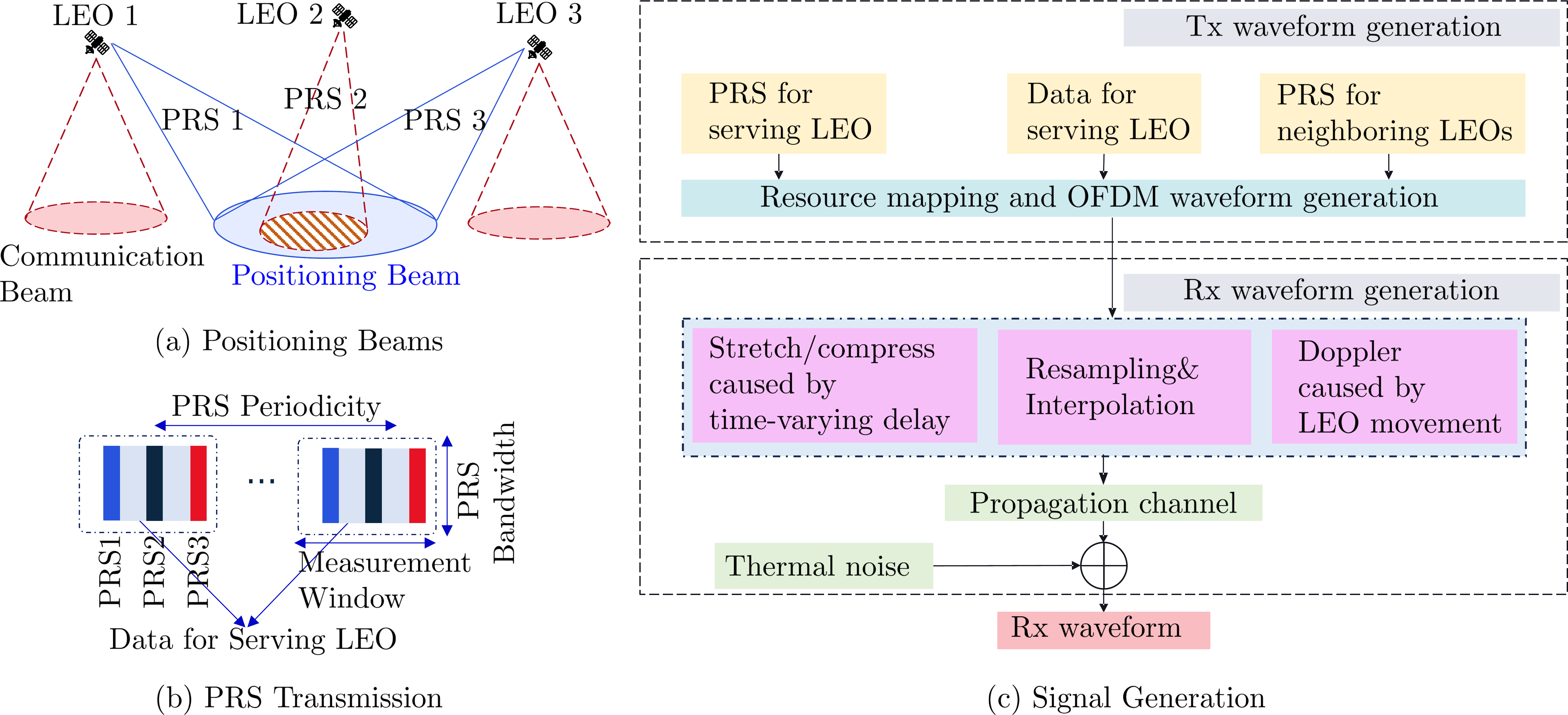}
    \captionsetup{font=small}
    \caption{Signal generation, and instances of transmission and reception (assuming 3 LEOs).}
    \label{fig:signal_generation}
    \vspace{-16pt} 
\end{figure*}
\section{Re-purposing NTN for Positioning}\label{sec:ntn_reuse_positioning} \nopagebreak
\subsection{Number of Visible Satellites}
To assess the feasibility of positioning with NTN, which is primarily designed for communication, we observe the number of visible satellites at different elevation angles. 
While a constellation may be designed to ensure 100\% global coverage with a specified minimum elevation angle for communication purposes--driven by factors such as link budget constraints or onboard processing limitations--the UE can still process positioning signals from satellites below this minimum elevation angle. Fig.~\ref{fig:ntn_pnt}b illustrates a typical LEO constellation with orbits at an altitude of 600 km and an inclination of 70$^\circ$, considering 30 orbits with 28 satellites in each orbital plane, resulting in 840 satellites in total. These parameters are chosen to reflect realistic LEO constellation designs that ensure global coverage and adequate satellite visibility~\cite{9840374,10009900}. 
The constellation is specifically designed for a minimum elevation angle of 30$^\circ$. Notably, even at the equator--where the spacing between orbits is largest--each UE sees at least two satellites. 
In Fig.~\ref{fig:ntn_pnt}c, we present the cumulative distribution function (CDF) of the number of visible satellites at above different minimum elevation angles, considering UEs dropped near the equator (worst case scenario). It is clear that the constellation illustrated in Fig.~\ref{fig:ntn_pnt}b will ensure 6-9 visible LEOs at a minimum elevation angle of 15$^\circ$, and this number will increase as the UE latitude increases. More visible satellites ensure sufficiently low geometric dilution of precision (GDOP) for positioning~\cite{dureppagari_ntn_10355106}. Note that the constellation depicted in Fig.~\ref{fig:ntn_pnt}b facilitates the visibility of both ascending (satellites moving toward the northern hemisphere, marked with red faces) and descending LEOs (satellites moving toward the southern hemisphere, marked with blue faces), ensuring improved GDOP. 

Despite many satellites being visible with a smaller minimum elevation angle, the number of satellites from which a receiver can \chb{successfully decode} radio signals may be smaller. This is due to the beamforming commonly used in LEOs deployed for communications, while GNSS is designed to provide broad coverage, allowing all receivers that can `see' the satellite to receive its signal. LEO satellites utilize a set of active narrow service beams illuminating a small area (typically 50 km diameter, corresponding to 3 dB beamwidth~\cite{3gpp::38821}) around the nadir at a given time (illustrated as narrow red beams in Fig.~\ref{fig:ntn_pnt}a). The satellite may have broader beams away from the nadir, consequently causing interference to the service beam of another satellite. 
However, these broader beams, henceforth called {\em positioning beams}, need to accommodate PRS transmission so that the UE can detect the neighboring satellites (illustrated as broader blue beams in Fig.~\ref{fig:ntn_pnt}a). Hence, to enable NTN positioning, the satellite network operator (SNO) must occasionally turn on the positioning beams to send PRS. \chb{Since the transmit power of the satellite is limited and the PRS sequence length limits the processing gain, the positioning beams directed toward different areas may be activated in a time-division multiplexed (TDM) pattern, allowing the satellite to allocate more power to transmit PRS.}
\subsection{Signaling from Network to UE}\label{sec:signal_network2ue}
\chb{Unlike TNs, where PRS is transmitted via DL shared channel resources and is only accessible to UEs registered with the network (i.e., in the RRC\_Connected state),  NTN connectivity requires PRS and assistance information to be available to UEs even in the RRC\_IDLE state. This is essential to enable standalone NTN positioning without relying on GNSS.} To acquire PRS from satellites, the UE must receive a navigation message (NAV MSG) containing critical positioning information, including satellite ephemeris and PRS details (e.g., sequence, timing, and frequency resources). This NAV MSG could be broadcast via a System Information Block (SIB) in NR-NTN or transmitted over a new physical DL channel. 
To ensure seamless integration of communication and positioning with minimal impact on the NTN UE, the PRS from each satellite should be transmitted over a few OFDM symbols and repeated at a specified periodicity. \chb{This PRS periodicity must align with the TDM pattern of the positioning beams, with gaps between transmissions to avoid inter-satellite interference caused by large differential delays.} For effective positioning, satellites in the NTN need to be tightly synchronized. The SNO should schedule PRS transmissions over the positioning beams of a subset of satellites, ensuring that the UE receives PRS signals from multiple satellites within a given measurement window.

To illustrate this, Fig.~\ref{fig:signal_generation}a presents a scenario involving three LEOs, where LEO 2 acts as the serving LEO, delivering communication and positioning services to UEs within its coverage area (highlighted with red stripes), while the non-serving LEOs direct their positioning beams towards the same coverage area of interest. Similar to communication beams, these positioning beams are assumed to be earth-fixed beams~\cite{3gpp::38821,3gpp::38811}, covering fixed ground areas and following a predetermined schedule, sweeping across different regions. Fig.~\ref{fig:signal_generation}b provides a snapshot of PRS transmission, illustrating PRS periodicity, the measurement window to detect PRS from multiple satellites, and the TDM pattern of the positioning beams from different satellites. During this measurement window, the UE conducts a blind search for PRS in the time domain, employing multiple time-frequency hypotheses. 
Hence, to minimize UE complexity, the measurement window should be kept small (ideally a few slots). It is important to note that when the PRS and data from the serving LEO are transmitted in the same frequency band, data transmission is muted during the PRS measurement window. This ensures that there is no interference from communication signals during PRS reception. 

\section{LEO-based NTN-positioning: Evaluation Framework}
A critical step in evaluating the performance of LEO-based positioning is the development of a robust simulation framework that accurately reflects key system components: (1) sampling satellites from realistic LEO orbits, (2) incorporating communication/positioning beam patterns, and (3) received signal generation and processing. 
Our framework considers the NTN channel model with large and small-scale fading based on 3GPP recommendations~\cite{3gpp::38811} and realistic Rx architecture. 
\subsection{Signal Generation}
Fig.~\ref{fig:signal_generation}c provides a detailed overview of the signal generation framework based on the NR-NTN waveform, comprising both Tx and Rx waveform generation. In the Tx waveform generation, the framework integrates PRS transmission from both serving and non-serving LEOs alongside data transmission for the serving LEOs, ensuring seamless communication and positioning. In LEO-based systems, Rx waveform generation is a critical component as it encapsulates the core characteristics of these systems. Specifically, in the Rx waveform generation, we account for the stretching and compression effects in the received signals due to time-varying delays, resampling and interpolation to capture these delays at the sample level considering the Rx sampling rate, and the Doppler shifts introduced by LEO movement-factors that are fundamental to LEO-based systems. To model large-scale and small-scale channel fading parameters, satellite Tx characteristics such as Tx power, antenna gains, and propagation losses, we use the radio channel model for NTN introduced  in~\cite{3gpp::38821,3gpp::38811}.
\vspace{-3mm}
\subsection{Receiver Architecture
}\label{sec:rx_architecture_and_results}

\begin{figure}[t]
    \centering
    \includegraphics[width=0.95\columnwidth]{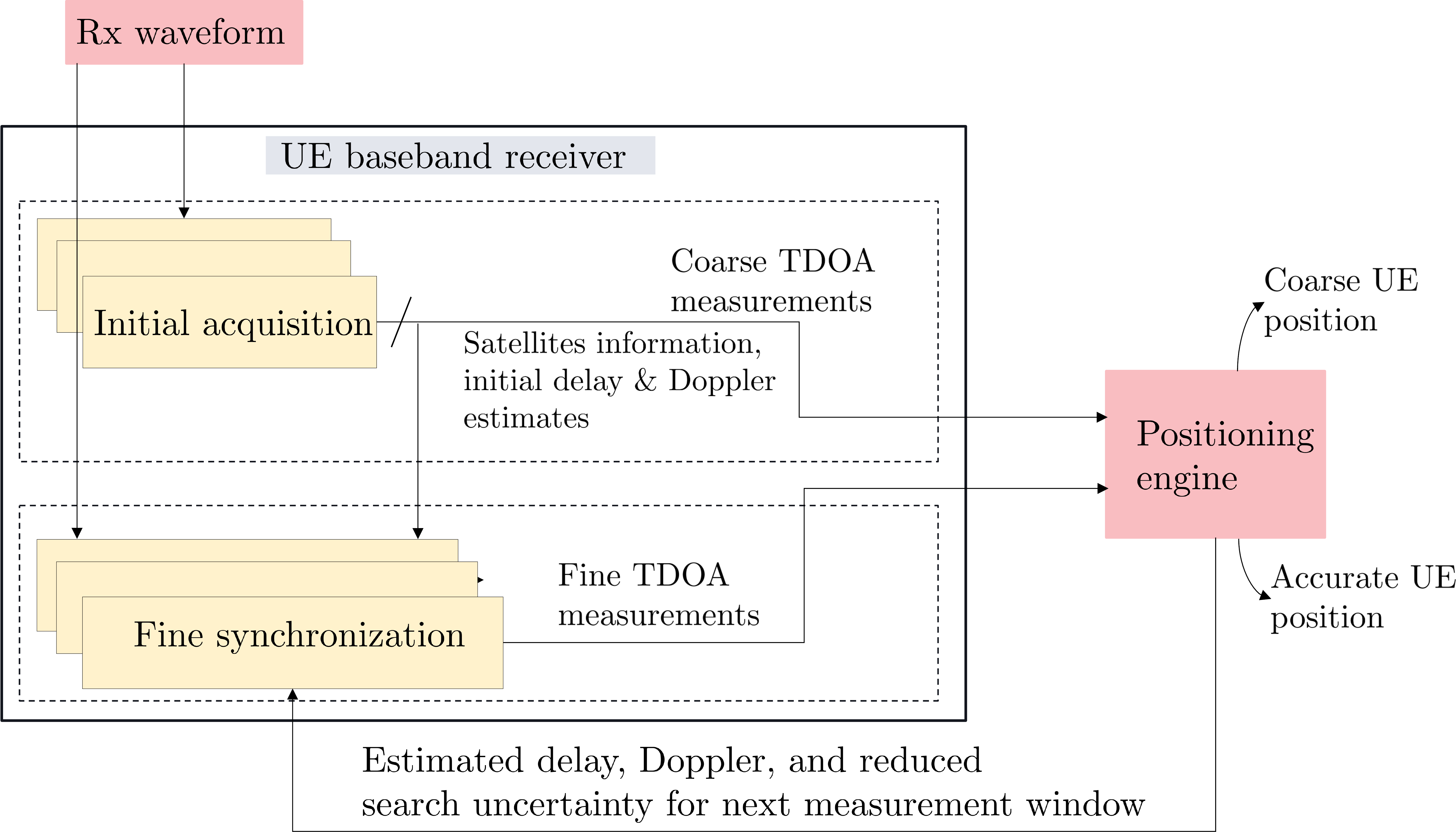}
    \captionsetup{font=small}
    \caption{Receiver block diagram capturing baseband processing at UE.}
    \label{fig:rx_blk_diagram}
    \vspace{-16pt} 
\end{figure}

The receiver architecture is outlined in Fig.~\ref{fig:rx_blk_diagram}. The receiver at the UE is broadly categorized into two main blocks, namely initial acquisition and measurement combining. Upon receiving the NAV MSG, the UE first aims to detect PRS signals from multiple satellites, acquiring initial delay (coarse TDOA estimates) and Doppler estimates. These initial measurements are then passed to the positioning engine for a coarse location estimate. This positioning engine can be a location server within the network for UE-assisted positioning or a localization application residing within the UE for UE-based positioning. 
Following the initial acquisition, the UE periodically estimates delay and Doppler offsets from multiple satellites, and these measurements are combined over time in the positioning engine to further improve positioning accuracy. 

Before getting into the details of Rx architecture components and the corresponding evaluation results, we first discuss the evaluation setup. Our evaluation considers the LEO constellation detailed in Section~\ref{sec:ntn_reuse_positioning} with UEs assumed to be uniformly distributed near the equator in a circular coverage area with a diameter of approximately 50 km, which is standard in 3GPP NTN evaluations~\cite{3gpp::38821,3gpp::38811}. Note that the  UE drop near the equator represents the {\em worst-case scenario} regarding LEO visibility and coverage. 
\chb{For PRS, subcarriers of an OFDM symbol (at 15 kHz subcarrier spacing (SCS)) are loaded with a pseudo-noise (PN) Gold sequence, as specified in~\cite{3gpp::38211}. This choice of 15 kHz SCS aligns with the standard NR-PRS configuration and ensures compatibility with sub-6 GHz NR UEs, though our framework can be extended to other NR numerologies, such as 30 kHz. Additionally, each LEO transmits a distinct PRS sequence using a unique PRS ID and TDM across PRS transmissions employed to avoid inter-satellite interference.} The PRS periodicity was set to 40 ms. Equivalent Isotropic Radiated Power (EIRP) density is configured to 34 dBW/MHz. \chb{The system operates in S-band at a 2 GHz carrier frequency (n256)}.  We consider 1 MHz and 5 MHz bandwidths and the corresponding Rx sampling rates of 10.56 MHz and 53.76 MHz, respectively. 1 MHz was chosen as a baseline for the following reasons: 1) Global Positioning System (GPS) civilian code uses approximately 1 MHz bandwidth, and 2) we aim to demonstrate LEO-based NTN positioning performance with minimal resources. On the other hand, 5 MHz was chosen to showcase improved accuracy comparable to GNSS. We assume 1 and 2 Rx ports at the UE.
\subsection{Blind Search for Initial Acquisition} Upon receiving NAV MSG, UE aims to detect PRS from multiple satellites to jointly estimate initial delay and Doppler offsets. This is achieved by conducting a blind search employing time and frequency hypotheses. The blind search involves UE locally generating the PRS waveform, correlating it with the received signal under different frequency hypotheses \chb{to estimate the Doppler offset and achieve frequency synchronization}, followed by peak detection. Note that, in LEO-based systems, we observe very high Doppler shifts on the order of 30-40 KHz due to their high orbital speeds, emphasizing the need for Doppler estimation and correction. The threshold for peak detection is derived to achieve a false alarm probability of $0.001$ in the case of additive-white-Gaussian noise (AWGN).  

Fig.~\ref{fig:prob_of_detect_vs_snr} presents the probability of PRS detection over different SNRs, considering one Rx port at the UE and one symbol PRS with 1 MHz and 5 MHz bandwidths. As depicted, higher SNR and bandwidth (higher bandwidth improves UE processing gain) improve detection probability. Note that increasing bandwidth from 1 MHz to 5 MHz does not necessarily result in 5x improvement in PRS detection due to several factors such as: 1) PRS detection probability is a non-linear function and is not directly proportional to the processing gain at the UE, and 2) multipath fading becomes more pronounced at higher bandwidth, which can degrade detection performance. 
Specifically, for 1 MHz bandwidth, PRS detection is unreliable for SNRs near or below -1 dB. Next, we will discuss how multiple Rx ports at the UE can improve PRS detection.
\subsection{Design Insights}
{\em Multiple Rx Ports.} NR smartphones are typically equipped with two Rx antenna ports in the S-band. This additional hardware can be leveraged to enhance PRS detection and speed up initial acquisition by exploiting combining gain and diversity. This is achieved by independently correlating the received signals at each Rx port and then combining the outputs. The key question is how to combine these outputs effectively. Coherent combining typically offers the highest gain but requires phase tracking and compensation across combining blocks, increasing receiver complexity. It also requires a line-of-sight (LOS) channel and identical small-scale fading across Rx ports. Given these requirements, non-coherent combining is often preferred. In an AWGN scenario, non-coherent combining with two Rx ports provides a gain of approximately 1.4 dB, as opposed to the 3 dB gain offered by coherent combining.
\begin{figure}[t]
    \centering
    \resizebox{0.82\columnwidth}{!}{
        \begin{minipage}{\columnwidth} 
            \centering
             \begin{subfigure}[b]{\columnwidth} 
                 \centering
                 \small
                 \includegraphics[width=1.03\textwidth]{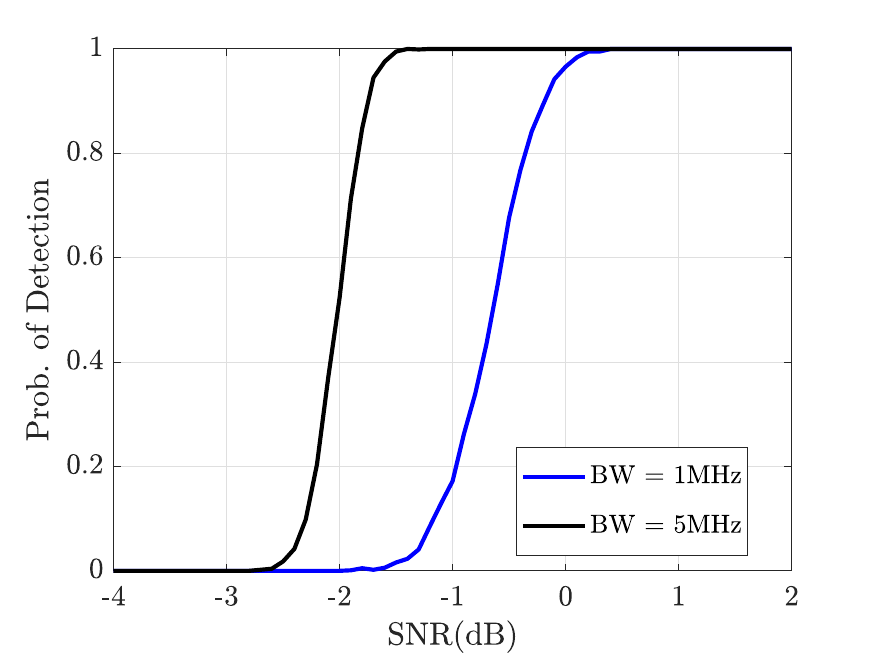}
                 \captionsetup{font=small}
                 \caption{Probability of Detection vs SNR.}
                 \label{fig:prob_of_detect_vs_snr}
             \end{subfigure}
            \hfill
             \begin{subfigure}[b]{\columnwidth} 
                 \centering
                 \small
                \includegraphics[width=0.94\textwidth]{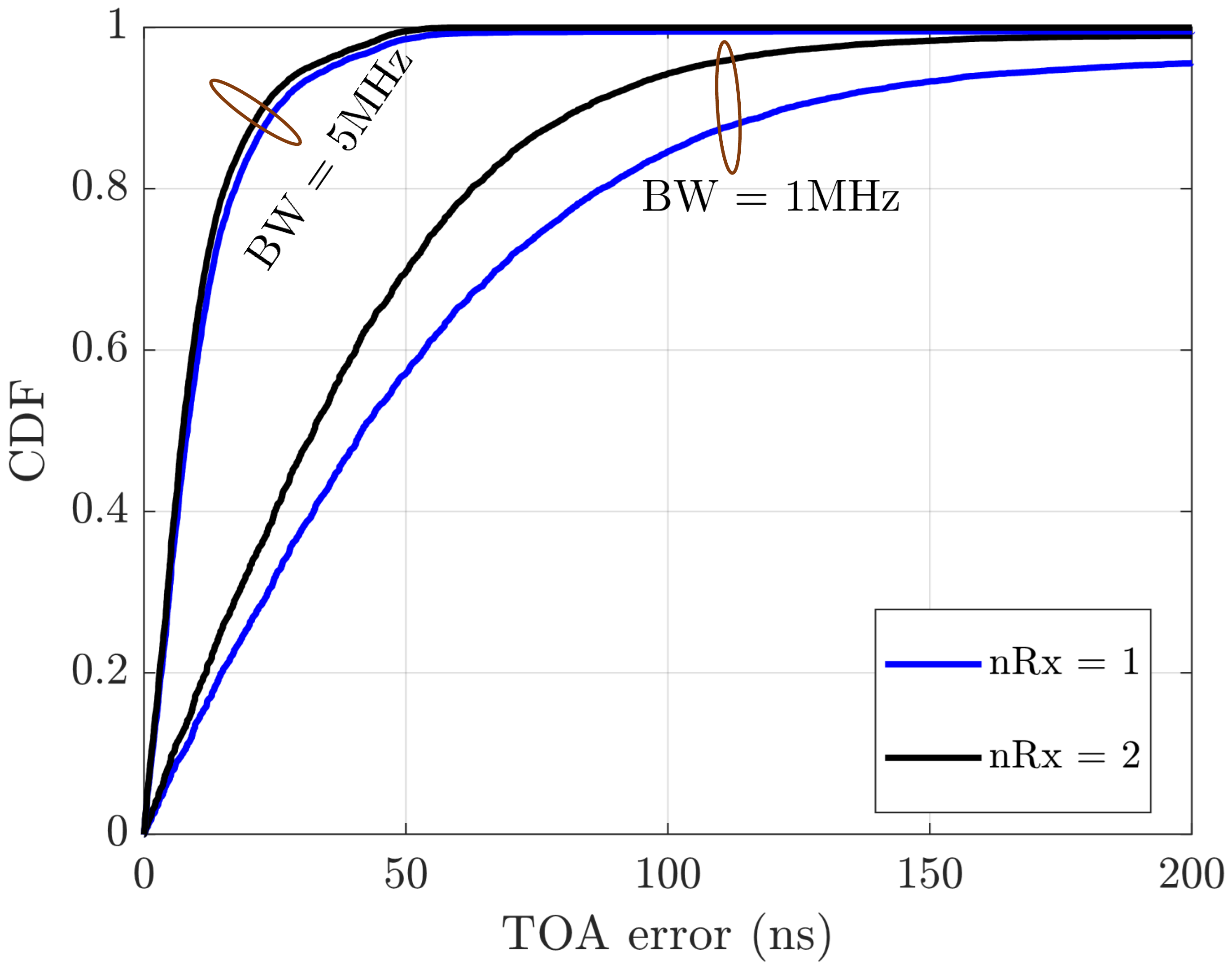}
                \captionsetup{font=small}
                \caption{CDF of TOA Error for 1 \& 5 MHz bandwidths.}
                \label{fig:cdf_toa_err_1mhz_5mhz}
             \end{subfigure}
        \end{minipage}
    }
     \captionsetup{font=small}
     \caption{{\em TOA error performance: }Probability of PRS detection for different SNRs and TOA error performance.}
     \label{fig:toa_err_performance}
     \vspace{-18pt} 
\end{figure}

We evaluated time-of-arrival (TOA) error performance using the discussed PRS detection mechanism with multiple Rx ports. Fig.~\ref{fig:cdf_toa_err_1mhz_5mhz} presents the CDF of TOA error for 1 MHz and 5 MHz bandwidths, considering one and two Rx ports at the UE. The CDF is plotted by compiling TOA error values across all detected satellites. It is evident that employing two Rx ports at the UE significantly improves TOA performance compared to using a single Rx port. This improvement is particularly pronounced in the 1 MHz scenario, where the lower bandwidth does not provide a high processing gain at the UE. 
Note that the CDF does not reach 1, specifically for the 1 MHz scenario, because some UEs fail to detect the PRS from certain satellites within the measurement window and are assigned very high TOA errors when the CDF is plotted.

\begin{figure}[t]
     \centering
     \captionsetup[subfigure]{font=small}
    \resizebox{0.88\columnwidth}{!}{ 
     \begin{subfigure}[b]{0.465\columnwidth}
         \centering
         \small
         \includegraphics[width=\textwidth]{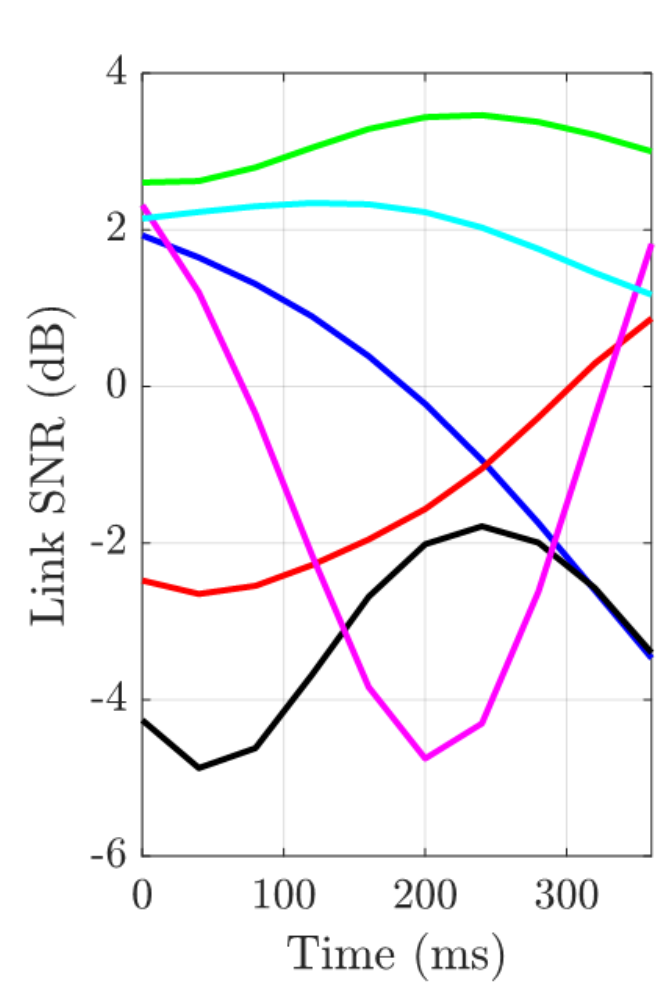}
        \captionsetup{font=small}
        \caption{SNR Evolution in Time.}
        \label{fig:snr_evolution}
     \end{subfigure}
     \hfill
     \begin{subfigure}[b]{0.505\columnwidth}
        \centering
        \small
         \includegraphics[width=\textwidth]{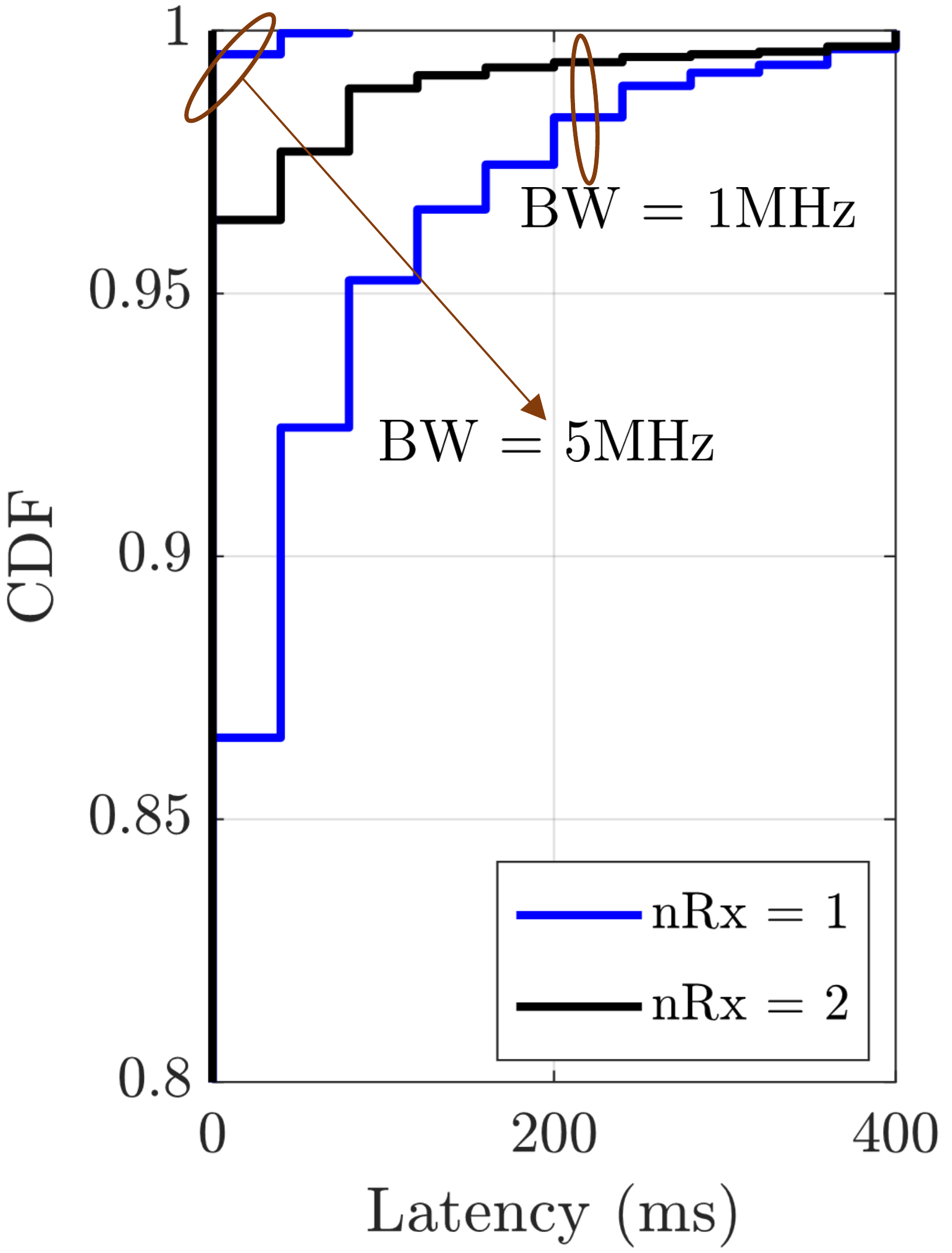}
         \captionsetup{font=small}
         \caption{CDF of TOA Latency.}
         \label{fig:cdf_toa_latency}
     \end{subfigure}
     }
     \captionsetup{font=small}
     \caption{{\em Searching over multiple PRS occasions: }(a) SNR evolution in time and the corresponding latency to acquire per satellite TOA considering 400 ms search window (1 \& 5 MHz bandwidth, 1 PRS symbol in 1 ms slot, and maximum of 10 occasions).}
     \vspace{-19pt} 
\end{figure}

{\em Searching PRS over Multiple Occasions.} It is essential to highlight that when operating bandwidth and SNR is low, the UE may not detect PRS on the first occasion, necessitating a search over multiple occasions. Notably, this may improve PRS detection probability as the channel evolves due to Doppler, facilitating UEs moving from a lower SNR regime to a higher SNR regime, as illustrated in Fig.~\ref{fig:snr_evolution} (SNR evolution for specific UEs with a non-serving LEO). However, searching over multiple occasions increases latency in acquiring TOA per satellite, thereby increasing positioning latency. Notably, employing multiple Rx ports at the UE strikes a balance, enhancing PRS detection probability while reducing latency. 

Fig.~\ref{fig:cdf_toa_latency} presents the CDF of TOA acquisition latency for 1 MHz and 5 MHz bandwidths (zoomed in between 0.8 and 1.0 for clarity), employing one and two Rx ports. For this evaluation, we assume a search window of 400 ms with a PRS periodicity of 40 ms, resulting in a maximum of 10 PRS occasions. Thus, the maximum latency in TOA acquisition is 400 ms, corresponding to no PRS detection. The CDF is plotted by compiling latency values for all detected satellites. As illustrated, searching PRS over multiple occasions significantly improves PRS detection. Specifically, for 1 MHz with one Rx port, the detection improved by 13.1\%. Moreover, employing multiple Rx ports at the UE and higher bandwidth further improves detection probability and reduces latency. Once coarse delay and Doppler estimates have been obtained, the computational complexity for tracking these estimates over time can be significantly reduced by performing peak detection with smaller and finer delay and Doppler hypotheses, typically done in the frequency domain.

{\em Measurement Combining.} Following the initial acquisition, TDOA measurements are combined across time to estimate UE location accurately. We employed a window-based combining method to accumulate TDOA measurements over time. For instance, with a PRS periodicity of 40 ms and a combining window of 400 ms (typically an integer multiple of the PRS periodicity), we collect up to 10 sets of positioning measurements to perform UE localization. During initial acquisition, particularly when an insufficient number of PRS signals are detected, the UE may need to search for PRS signals across multiple occasions. The measurements obtained during these searches are combined to derive a coarse location estimate of the UE. Once the initial acquisition is achieved, the measurements combined within the combining window help further refine positioning accuracy, particularly in scenarios with low bandwidth and low SNR. \chb{Importantly, we do not assume constant propagation delay across the measurement window.}

It is crucial to recognize that the set of LEOs visible to a particular UE may change over time, causing variations in the acquired positioning measurements. Additionally, as the channel conditions evolve, the UE may occasionally fail to acquire measurements from certain LEOs, even when a sufficient number of LEOs remain visible. This behavior results in a series of positioning measurements with different TDOA patterns. For example, consider a scenario where a UE acquires PRSs from LEOs 1 and 2 during the first measurement occasion, yielding one TDOA measurement. Subsequently, the UE might obtain two TDOA measurements from LEOs 1, 2, and 4 during the second occasion and two more from LEOs 2, 3, and 4 during the third occasion. \chb{These varying sets of TDOA measurements, along with the satellite positions at the PRS transmission times, are then provided to the positioning engine, which employs weighted nonlinear least squares (WNLS)~\cite{1130282269234990848} to calculate the UE location.} As time progresses, the combining window advances by one PRS occasion to gather new TDOA measurements. 
\begin{figure}[t]
    \centering
    \resizebox{0.82\columnwidth}{!}{
        \begin{minipage}{\columnwidth} 
            \centering
             \begin{subfigure}[b]{\columnwidth}
                 \centering
                 \small
                \includegraphics[clip, trim=0.2cm 0cm 0cm 0cm,width=0.95\columnwidth]{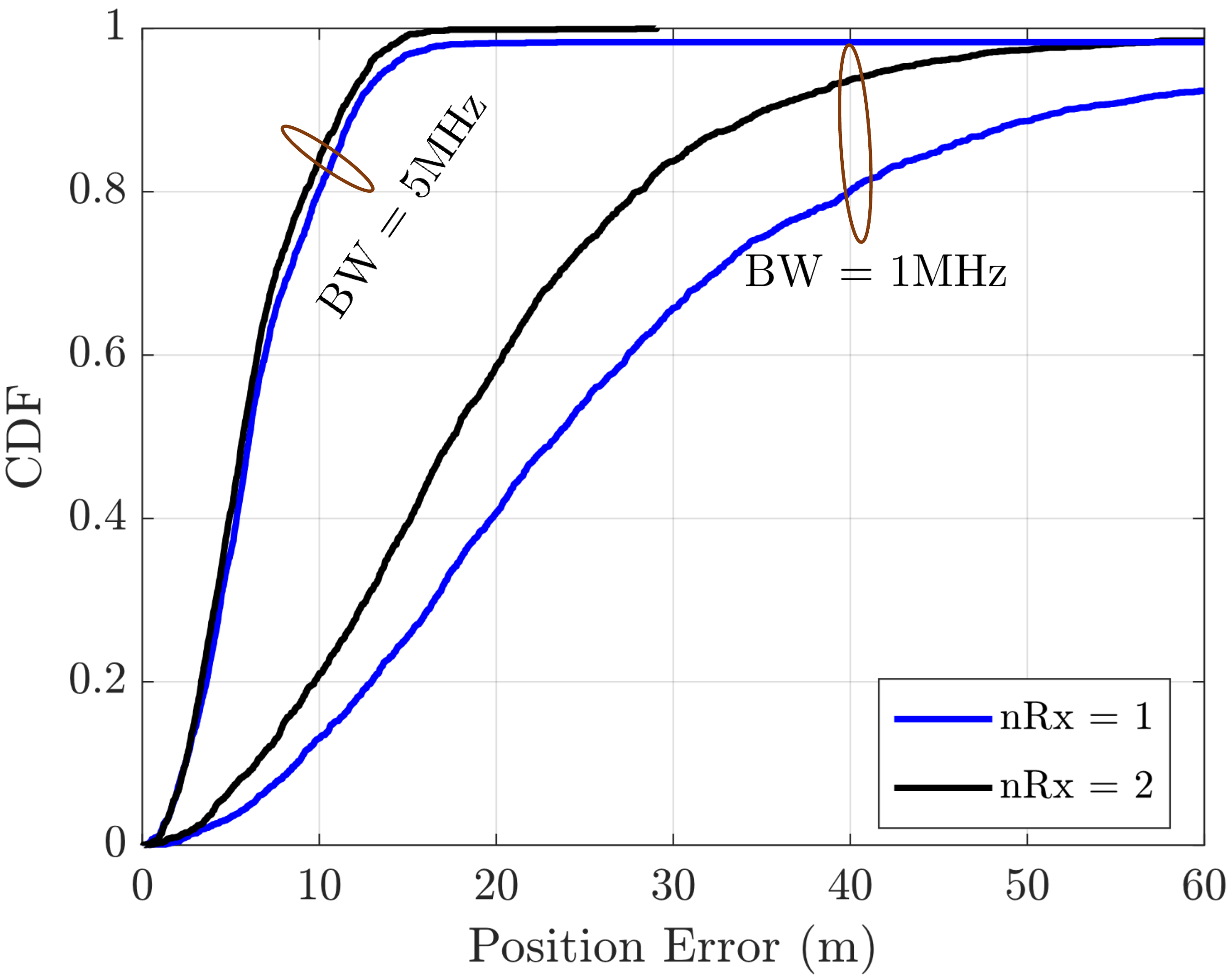}
                \captionsetup{font=small}
                \caption{Single PRS Occasion.}
                \label{fig:cdf_pos_err_one_shot}
             \end{subfigure}
            \hfill
             \begin{subfigure}[b]{\columnwidth}
                 \centering
                 \small
                \includegraphics[clip, trim=0cm 0cm 1.0cm 2.0cm, width=1.03\columnwidth]{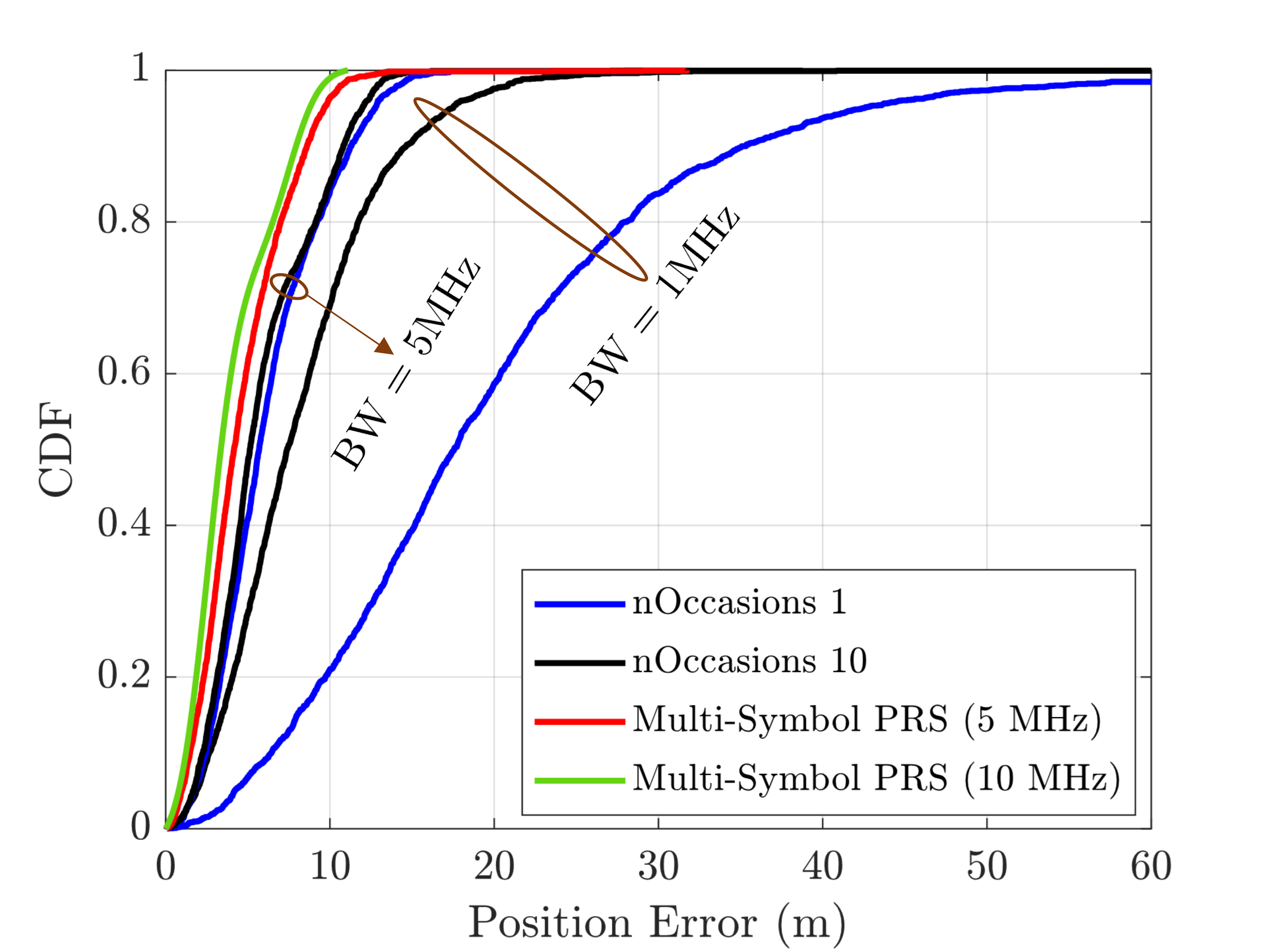}
                \captionsetup{font=small}
                \caption{Measurement Combining and Multi-Symbol PRS}
                \label{fig:cdf_pos_err_combining_multiprs_1mhz_5mhz}
             \end{subfigure}
        \end{minipage}
    }
    \captionsetup{font=small}
    \caption{\chb{{\em Positioning performance:} CDF of position error using 1 symbol PRS in a 1 ms slot: (a) one PRS occasion, (b) combining window of 400 ms for 1 MHz and 5 MHz with varying number of occasions [1 10] (nOccasions). Multi-symbol PRS in (b) considers 14 PRS symbols in one PRS occasion in a 1 ms slot for 5 MHz and 10 MHz bandwidths.}}
    \label{fig:pos_perf_one_shot_tracking}
     \vspace{-20pt} 
\end{figure}

\chb{We now evaluate positioning performance for three scenarios: a single PRS occasion, measurement combining across multiple PRS occasions, and multi-symbol PRS in a single occasion. To reflect realistic scenarios, we assume a maximum of four visible satellites transmitting PRS, as LEO constellations are primarily designed for communication rather than dedicated positioning. 
This results in a higher GDOP and consequently limits the achievable positioning accuracy. Fig.~\ref{fig:cdf_pos_err_one_shot} illustrates the positioning accuracy for a single PRS occasion, 
where one PRS symbol is transmitted within a 1 ms slot. The evaluation considers two bandwidths (1 MHz and 5 MHz) and two Rx port configurations (1 and 2 Rx ports, referred to as nRx in the plots). Consistent with TOA error trends, employing two Rx ports enhances positioning accuracy, with a more pronounced effect in the 1 MHz case due to its lower processing gain. Specifically, for 1 MHz bandwidth, using two Rx ports improves positioning accuracy by 25\% compared to one Rx port, while for 5 MHz bandwidth, the improvement is 6\%. Additionally, increasing the bandwidth from 1 MHz to 5 MHz with one Rx port enhances positioning accuracy by 74\%, whereas the improvement is 68\% with two Rx ports, underscoring the importance of bandwidth for improved positioning accuracy.}

\chb{Fig.~\ref{fig:cdf_pos_err_combining_multiprs_1mhz_5mhz} presents positioning accuracy when measurements are combined across multiple PRS occasions. For this evaluation, we assume a combining window of 400 ms with a PRS periodicity of 40 ms, allowing the UE to combine measurements from up to 10 PRS occasions (referred to as nOccasions in the plots). The 400 ms window was chosen to improve PRS detection and measurement availability while maintaining manageable receiver complexity. The positioning error CDFs compare the performance of a single PRS occasion and measurement combining across 10 PRS occasions. The results indicate that measurement combining improves positioning accuracy significantly for 1 MHz but not prominently for 5 MHz.}

{\em Multi-Symbol PRS. }Although coarse positioning accuracy achieved with a single-symbol PRS in a 1 ms slot may suffice for initial NTN access and UL synchronization without GNSS assistance, further accuracy improvements are essential for application-level location services such as navigation. \chb{This can be achieved by transmitting multiple PRS symbols over an extended duration, similar to GNSS. Fig.~\ref{fig:cdf_pos_err_combining_multiprs_1mhz_5mhz} demonstrates that multi-symbol PRS (14 PRS symbols over 1 ms for 5 MHz and 10 MHz bandwidths) significantly enhances positioning accuracy. Specifically, with 10 MHz bandwidth, we achieve sub-10m accuracy for close to 99\% of UEs, a performance comparable to GNSS~\cite{Edkgps2005}.} This improvement arises from the longer sequence duration, which enhances autocorrelation properties and noise immunity and reduces cross-correlation. These results are particularly notable given that our assumptions limit the maximum number of LEOs transmitting PRS to four, emphasizing the benefits of an improved link budget. Furthermore, higher NTN bandwidths further improve accuracy, reinforcing the potential of LEO-based NTN positioning as a strong complement to GNSS.

Unlike in TN, PRS from multiple satellites frequency-division multiplexed (FDMed) in a comb pattern does not guarantee orthogonality at the UE because of the large delay and frequency difference among satellites. On the other hand, spreading the LEO PRSs over time (like the Wideband Code Division Multiple Access (WCDMA) signal used in GNSS) may not be ideal due to the {\em near-far problem} of LEOs. Specifically, the SNRs of the received PRSs from the nearest and farthest LEOs can differ by more than 12 dB. Hence, careful resource planning (e.g., TDM with gaps) is required to ensure no interference from the PRS of other satellites.

{\em Generality of Evaluation Framework. } 
Our evaluation framework is flexible and adaptable, accommodating different constellation configurations and UE distributions. While this study uses a reference constellation outlined in Section~\ref{sec:ntn_reuse_positioning}, the framework allows modifying key parameters, such as inclination, minimum elevation angle, orbital planes, and satellite density, to analyze alternative constellation designs. Also, note that the 840-satellite constellation used in this work is a conservative representation compared to anticipated larger deployments. Additionally, we assume a maximum of four satellites transmitting PRS for a given coverage area, which is a modest assumption compared to future LEO deployments and prioritization of positioning services. The framework also supports diverse UE distributions across various geographic regions and latitudes.  In this work, we focus on UE distribution near the equator, representing a worst-case scenario due to wider orbital plane spacing resulting in fewer visible satellites. At higher latitudes, increased satellite visibility provides more flexibility for PRS scheduling, and yields improved positioning accuracy. Consequently, the accuracy numbers reported in this study are expected to improve under denser constellations and UEs at higher latitudes.

{\em GNSS vs. LEO-based NTN Positioning.} There are a few fundamental differences between GNSS and LEO-based NTN positioning systems. First, GNSS satellites continuously transmit positioning signals enabling the receiver to coherently combine and achieve high processing gains~\cite{Edkgps2005} (on the order of 43 dB with sequence spread over 1 ms). In contrast, in NR-NTN, the processing gain is limited by the PRS sequence length. 
Second, GNSS receivers can keep combining the received samples continuously thanks to the {\em always-on} transmission, whereas, in NR-NTN, intermittent PRS transmission (multiplexed with communication symbols) limits combining gain. Third, exploiting phase coherence, GNSS receivers employ advanced positioning methods such as carrier phase tracking, which can provide cm-level accuracy~\cite{10009900}. In contrast, LEO-based systems may not guarantee carrier phase coherence or may not continuously transmit a reference signal over which the carrier can be measured. Fourth, there may be a significant SNR imbalance from neighboring LEOs because of the near-far problem, which is less pronounced in GNSS. \chb{While isoflux antennas can mitigate this effect in GNSS by shaping gain across the footprint, such designs are not commonly used in communication-focused LEO constellations, making SNR imbalance a practical concern for NTN positioning.}
\vspace{-2mm}
\section{Concluding Remarks and Future Work}
A key feature of NTN for 6G is expected to be an addition of multi-LEO satellite-based PNT capability of UEs with different accuracy regimes targeted for a wide range of applications (e.g., obtaining UL synchronization, localization for registration, emergency calls, regulations, precise location services). This study provides comprehensive design insights for reusing NR-NTN for positioning. However, several key areas warrant further investigation

{\em NTN Positioning as an Additional Service.} As NR-NTN continues to evolve, there is an opportunity to provide positioning as an additional service offered by SNOs delivering cellular services over NTN. Unlike TNs, NTN systems are not currently mandated to provide location services. However, in the future, SNOs may need to build location services into their NTN offerings to meet regulatory requirements and manage location-based content delivery and subscriptions. Our work lays the initial design guidelines to that end.

{\em Addressing UE Clock Drift Issues. }In this study, we did not account for UE clock drift. This drift restricts the length of the measurement window (as shown in Fig.~\ref{fig:signal_generation}c) and consequently limits the separation between PRS occasions from multiple satellites, over which the UE can acquire one set of TDOA measurements. Future studies may investigate methods to mitigate the effects of UE clock drift, potentially extending the measurement window and enhancing positioning accuracy.

{\em Handling Satellite Clock Drift and Synchronization.} 
For this study, we assumed that DL frame boundaries are perfectly synchronized across satellites, which is not necessarily required for communication but is crucial for accurate positioning. Maintaining network-level synchronization may impose additional overhead for SNOs. Future work could investigate the long-term stability of LEO satellite clocks, the impact of clock drift, and the need to communicate clock offsets to UEs in case of inaccuracies.

{\em Exploring UL Positioning.} 
This study primarily focused on DL-only positioning based on DL PRS. In addition, future research may explore UL positioning based on Sounding Reference Signal (SRS) to enhance positioning accuracy further. 
\vspace{-5mm}
\bibliographystyle{IEEEtran}
\bibliography{hokie}
\end{document}

%% file: notation.tex
\def\nb0{{\mathbf{0}}}
\def\nb1{{\mathbf{1}}}

